\documentstyle[12pt,epsf,amssymb]{article}
\def\dfrac#1#2{{\displaystyle {#1 \over #2}}}

\def\simge{\mathrel{\rlap{\raise 0.511ex \hbox{$>$}}{\lower 0.511ex \hbox{$\sim$}}}}
\def\simle{\mathrel{\rlap{\raise 0.511ex \hbox{$<$}}{\lower 0.511ex \hbox{$\sim$}}}} 
\def\slash#1{\setbox0=\hbox{$#1$}\dimen0=\wd0                     \setbox1=\hbox{/} \dimen1=\wd1 \ifdim\dimen0>\dimen1                    \rlap{\hbox to \dimen0{\hfil/\hfil}} #1                        \else                                       
      \rlap{\hbox to \dimen1{\hfil$#1$\hfil}}  
      /   \fi}                                         
\newcommand{\be}{\begin{equation}}
\newcommand{\ee}{\end{equation}}
\newcommand{\bea}{\begin{eqnarray}}
\newcommand{\eea}{\end{eqnarray}}
\newcommand{\msb}{\overline{\rm{MS}}}
\newcommand{\mev}{\,{\rm MeV}}   
\newcommand{\gev}{\,{\rm GeV}}   
\newcommand{\mqri}{m^{{\scriptsize{\rm RI}}}_q}
\newcommand{\mri}{m^{{\scriptsize{\rm RI}}}}
\newcommand{\ri}{{\scriptsize{\rm RI}}}
\newcommand{\mlri}{m^{{\scriptsize{\rm RI}}}_\ell}
\newcommand{\msri}{m^{{\scriptsize{\rm RI}}}_s}
\newcommand{\mlmsbar}{m^{\scriptsize \overline{\rm MS}}_\ell}
\newcommand{\msmsbar}{m^{\scriptsize \overline{\rm MS}}_s}

\newcommand{\mqmsbar}{m^{\scriptsize \overline{\rm MS}}}
\newcommand{\mlrgi}{m^{\mbox{\scriptsize{\rm RGI}}}_\ell}
\newcommand{\msrgi}{m^{\mbox{\scriptsize{\rm RGI}}}_s}
\newcommand{\mqrgi}{m^{\mbox{\scriptsize{\rm RGI}}}_q}

\newcommand{\Oa}{{\cal O}(a)}

\newcommand{\Oaa}{{\cal O}(a^2)}
\newcommand{\suno}{\widehat \sigma _1}
\newcommand{\sdue}{\widehat \sigma _2}

\newcommand{\Dslash}{\slash D}
\newcommand{\dslash}{\slash \partial}
\newcommand{\cqp}{c'_q}
\newcommand{\cngi}{c_{\scriptstyle{\rm NGI}}}

\def\nabstar#1{\nabla\kern-0.5pt\smash{\raise 4.5pt\hbox{$\ast$}}                \kern-4.5pt_{#1}}

\def\drvstar#1{\partial\kern-0.5pt\smash{\raise                                  4.5pt\hbox{$\ast$}} \kern-5.0pt_{#1}}

\def\lvec#1{\setbox0=\hbox{$#1$}
    \setbox1=\hbox{$\scriptstyle\leftarrow$} #1\kern-\wd0\smash{
    \raise\ht0\hbox{$\raise1pt\hbox{$\scriptstyle\leftarrow$}$}}
    \kern-\wd1\kern\wd0}
\def\lrvec#1{\setbox0=\hbox{$#1$}
    \setbox1=\hbox{$\scriptstyle\leftrightarrow$}
    #1\kern-\wd0\smash{
\raise\ht0\hbox{$\raise1pt\hbox{$\scriptstyle\leftrightarrow$}$}}
    \kern-\wd1\kern\wd0}
\def\rvec#1{\setbox0=\hbox{$#1$}
    \setbox1=\hbox{$\scriptstyle\rightarrow$}
    #1\kern-\wd0\smash{
    \raise\ht0\hbox{$\raise1pt\hbox{$\scriptstyle\rightarrow$}$}}
    \kern-\wd1\kern\wd0}

\begin{document}

\begin{titlepage}
\begin{flushright}
FTUV 99/22\\
IFIC 99/24 \\
RM3-TH/99-4 \\
Roma 1260/99
\end{flushright}
\vskip 2.4cm
\begin{center}
{\Large \bf Light Quark Masses from  Lattice Quark\\ 
Propagators at Large Momenta}
\vskip1.3cm 
{\large\bf D.~Becirevic$^a$, V.~Gim\'enez$^b$, V.~Lubicz$^c$ and 
G.~Martinelli$^a$}\\

\vspace{1.cm}
{\normalsize {\sl 
$^a$ Dip. di Fisica, Univ. di Roma ``La Sapienza" and INFN,
Sezione di Roma,\\ P.le A. Moro 2, I-00185 Rome, Italy\ ,\\
\vspace{.25cm}
$^b$ Dep. de Fisica Teorica and IFIC, Univ. de Valencia,\\
Dr. Moliner 50, E-46100, Burjassot, Valencia, Spain\ ,\\
\vspace{.25cm}
$^c$ Dip. di Fisica, Univ. di Roma Tre and INFN,
Sezione di Roma Tre, \\
Via della Vasca Navale 84, I-00146 Rome, Italy\ .}\\
\vskip1.cm
{\large\bf Abstract:\\[10pt]} \parbox[t]{\textwidth}{ 
We compute non-perturbatively the average up-down and 
strange quark masses from the large momentum (short-distance) behaviour of the quark propagator in the Landau gauge. This
 method, which has never been applied so far, does not require the explicit calculation of the quark mass 
renormalization constant. Calculations were performed in 
the quenched  approximation, by using  $\Oa$-improved  Wilson fermions. The main results of this study are 
$\mlri(2 \gev)=5.8(6)$~MeV and $\msri(2 \gev)=136(11)$~MeV.
Using the relations between different schemes, obtained from the
available four-loop anomalous dimensions, we also find  
 $\mlrgi=7.6(8)$~MeV and $\msrgi=177(14)$~MeV, and the $\msb$ masses,  $\mlmsbar(2 \gev) =4.8(5)$~MeV and $\msmsbar(2 \gev) =111(9)$~MeV. }}
\end{center}
\vspace*{1.cm} 
{\footnotesize{\tt PACS numbers: 12.15Ff,\ 11.15.Ha,\ 12.38.Gc.}}

\end{titlepage}

\setcounter{footnote}{0}
\setcounter{equation}{0}

\section{Introduction}
\label{sec:intro}

\par
Determination of quark masses is becoming one of
the most intensive field of investigation in lattice
QCD~\cite{mq_ape94}--\cite{mq_dwall}. The accuracy of the predictions is
significantly improving mainly because of two recent theoretical
developments:
\begin{itemize}
\item  Non-perturbative renormalization procedures have 
been introduced~\cite{npm,npm_sf} in order to remove the systematic 
uncertainties coming from  the truncation of perturbative series in
calculation of the relevant renormalization
 constants. These procedures also provide an appropriate non-perturbative, 
 short distance definition of the quark masses either in the so-called RI-MOM 
 or Schr\"odinger functional schemes. The relation 
 between the mass in the RI-MOM scheme (which will be used in this study) 
and in the $\msb$ scheme~\footnote{ 
The conversion of the results to the $\msb$ scheme is only necessary 
for comparison with other calculations for which the quark masses are 
renormalized perturbatively. Otherwise, the method of Ref.~\cite{npm} 
allows, in principle, to obtain the renormalized quark masses for the 
RI-MOM scheme in a completely non-perturbative way.}, 
or the renormalization group invariant mass, is known at 
next-to-next-to-leading order (${\rm N^2LO}$)~\cite{francolub}, and very recently even at
${\rm N^3LO}$~\cite{chetyrkin}, in continuum perturbation theory. 

\item A second important  theoretical progress is the reduction of 
finite cut-off ($\Oa$) effects obtained 
by improving the lattice fermion action and operators. The
perturbative procedure, 
proposed in Refs.~\cite{sw,heatlie}, has been 
recently extended to a fully non-perturbative $\Oa$-improvement by 
{\sc Alpha} collaboration~\cite{a1}, so that the remaining
discretization errors are only of ${\cal O}(a^2)$. 
\end{itemize}
\noindent
In the past year, 
several independent lattice determinations of the light quark 
masses~\cite{mq_noi}--\cite{qcdsf} have been presented, adopting both
non-perturbative renormalization procedures and  non-perturbative improvement.

\par
The two standard definitions of the lattice  quark masses 
are based on  the vector (VWI) and  the axial-vector (AWI) chiral 
Ward identities~\cite{bochicchio}. The VWI relates the bare quark 
mass to the value of the Wilson hopping parameter, $2 am = (1/\kappa - 
1/\kappa_{crit})$. With this definition, one can easily show that the
mass renormalization constant is $Z_m(\mu)
=Z_S^{-1}(\mu)$. The definition based on the AWI 
is $2 a\bar m = \langle \alpha 
\vert \partial_\mu A_\mu \vert \beta \rangle/ \langle \alpha \vert \,P \vert
\beta \rangle$, where $\partial_\mu A_\mu$ and $P$ are the divergence of the 
(improved) axial vector current and the pseudoscalar density, 
respectively. In this case, $Z_{\bar m}(\mu)= Z_A/Z_P(\mu)$.

\par
In this paper, in order to calculate the renormalized quark mass, 
we adopt a new method based on the study of the large-$p^2$ behaviour of 
the renormalized quark propagator. The method is based on the idea 
that at large Euclidean momenta it is possible to match lattice and 
continuum correlators by requiring the vanishing of chirality violating form 
factors~\cite{unpub,latt98}. This procedure is justified by the following
two observations. The first is that at large momenta the renormalized
perturbation theory becomes chirally invariant (explicit chiral symmetry 
breaking effects induced by the regularization are reabsorbed by imposing 
the validity of the chiral Ward identities, while violations from the 
non-vanishing quark masses, disappear at large momenta). The second
observation is that the contributions due to the spontaneous breaking 
of chiral symmetry, which are absent in perturbation theory, die off at
large momenta.
Thus, both effects decrease as we go deeper into the Euclidean region.

\par
The simplest application of this idea is the possibility of relating the quark mass to the renormalized quark propagator
\bea
\widehat S(p) = \frac{i \slash{p}}{p^2} \, \suno (p^2) + \frac{\sdue (p^2)}{p^2}
\, , \label{eq:sigma12}
\eea
since, at large $p^2$, we expect~\cite{plderaf}
\bea
 \sdue (p^2) \simeq  m  + \langle \bar q q \rangle \frac{4 \pi \alpha_s}{3  p^2} +
{\cal O} \bigl( 1/p^4 \bigr) \,.  \label{eq9} \eea
In particular, the quark  mass renormalized in the RI-MOM scheme, can be directly extracted from the quark propagator  renormalized in the same scheme by using
\bea
\mqri (\mu) = \frac{1}{12} \, {\rm Tr} \biggl[ \widehat S^{-1}(p; \mu) \biggr]_{p^2=\mu ^2}\quad ,
\label{eq:master}
\eea
where the trace is over both color and spin indices~\footnote{We note in passing, that by using this method we were not able to extract the value of the quark condensate~(\ref{eq9}).}.
$\widehat S(p; \mu)$ is the (improved) quark propagator renormalized
 at some scale $\mu$. Since the quark propagator is a gauge dependent 
quantity, the definition of the RI-MOM mass also depends on the gauge. 

\par At  large momenta and up to discretization errors,
Eq.~(\ref{eq:master}) is  equivalent to the definition of the quark mass
based on the AWI.  Chiral symmetry provides a relation between 
the inverse quark propagator, $ \widehat S(p; \mu)^{-1}$, and the
amputated Green function of the pseudoscalar density, $\widehat \Lambda_5(p;\mu)$, computed between external (off-shell) quark states of equal momenta $p$. The AWI then reads:
\be
2 \, \widehat m_q (\mu) \, \widehat \Lambda_5 (p; \mu) = \gamma_5 \, \widehat 
S^{-1}(p; \mu) + \widehat S^{-1}(p; \mu) \, \gamma_5 \quad .
\label{eq:wid}
\ee
All quantities in Eq.~(\ref{eq:wid}) are assumed to be renormalized 
(and improved) in the same scheme, and at the same scale $\mu$. In the
RI-MOM scheme
(and in a fixed gauge), the Green function $\widehat \Lambda_5 (p; \mu)$ satisfies the following renormalization condition~\cite{npm}:
\bea
\frac{1}{12} \, {\rm Tr} \biggl[ \gamma_5 \widehat \Lambda_5 (p; \mu) \biggr] 
_{p^2=\mu ^2} = 1\quad .
\label{eq:rcond}
\eea
By tracing both sides of Eq.~(\ref{eq:wid}) with $\gamma_5$ and 
by using Eq.~(\ref{eq:rcond}),  the relation~(\ref{eq:master}) is readily
derived. Note that if $\mu$ in Eq.~(\ref{eq:master}) is not chosen  in the
perturbative region, {\it i.e.} $\mu \gg \Lambda_{QCD}$, the
definition of the quark mass  will be affected by  non-perturbative, 
chirally-breaking contributions proportional to the quark condensate and
higher-dimensional operators, appearing in higher power corrections
($\propto 1/p^{2n}$).

\par
The advantage in determining the masses from the quark propagators is
that it is not necessary to calculate explicitly the mass renormalization
constants ({\it i.e.} $Z_S(\mu)$ or $Z_P(\mu)$). This is merely a
consequence of the  fact that the renormalized quark propagator is
directly expressed in terms  of the renormalized quark mass. Unlike in the
case of the VWI, the critical value of the hopping parameter,
$\kappa_{crit}$, is also not needed (which is the advantage inherent to
the use of the AWI). There is, however, 
one renormalization constant for  any quark-mass definition: in our case,
using Eq.~(\ref{eq:master}),  this  is  the quark-field renormalization
constant $Z_q$. In the RI-MOM scheme $Z_q$ is fixed by the following
renormalization condition:
\bea
\frac{i}{48} \, {\rm Tr} \left[ \gamma_\mu \frac{\partial \widehat 
 S^{-1}(p; \mu)}{\partial p_\mu} \right] _{p^2=\mu ^2} \ \equiv \
\frac{i}{48} \, Z_q(\mu) \, {\rm Tr} \left[ \gamma_\mu \frac{\partial 
 S^{-1}(p)}{\partial p_\mu} \right] _{p^2=\mu ^2} = 1\quad .
\label{eq:rcs}
\eea
In summary, Eqs.~(\ref{eq:master}) and (\ref{eq:rcs}) are all we need to
extract  quark masses from  propagators. The procedure becomes
rather complicated, however, if we want to extend it to the
non-perturbatively improved case.
The drawback with the method of Ref.~\cite{npm} is that the improvement program
(which was initially carried out for on-shell quantities) must be
extended to off-shell Green functions on non-gauge invariant states and 
involves  additional counterterms for a full ${\cal O}(a)$ 
improvement~\cite{unpub}. The strategy followed in this  case will be
illustrated in detail in Sec.~\ref{sec:improv}. 

\par
In this paper, Eqs.~(\ref{eq:master}) and (\ref{eq:rcs}) have been used to compute the average
up-down  and the strange quark masses, by performing a lattice QCD 
calculation in the quenched  approximation. 
We use the non-perturbatively improved  action~\cite{a1}, 
and improve the quark propagator in the chiral limit. The
$\Oa$-improvement procedure for off-shell (gauge non-invariant) 
quantities has been discussed in Ref.~\cite{unpub}. 
Since that paper has not  been published yet, we will describe here 
in some detail the specific case of the quark propagator. For technical 
reasons, which are related to the mixing with non gauge-invariant 
higher-dimensional operators (see below), we are not able to improve the 
propagator out of the chiral limit~\footnote{
The general problem of the improvement out of the chiral limit has not been
solved yet, although several interesting  proposals
exist~\cite{unpub,massimo,petronzio,steve}.}. 
Therefore, our determination of the quark masses is affected  by 
${\cal O}(g_0^2 am)$ systematic errors. Since the value 
of the inverse lattice spacing in this simulation is $a^{-1} \simeq 2.72$ GeV,
these errors are expected to be negligible for the strange and 
the light quark masses.

\par
We conclude this section by summarizing the main results of this paper.
From the study of the quark propagator, we extract the
(quenched) light  and strange quark masses in the RI-MOM scheme:
\bea
\label{RImasses}
\mlri (2\gev) = 5.8(6) \mev \quad , \quad \msri (2\gev) = 136(11) \mev \quad .
\eea 
\noindent
These results are in very good agreement with those of 
Ref.~\cite{mq_noi}, namely
$\msri (2 \gev ) = 138(15)$~MeV, and $\mlri (2 \gev)=5.6(5)$~MeV. 

\par
Using the N$^3$LO perturbative formulae of Ref.~\cite{chetyrkin}, we obtain
the renormalization group invariant quark masses 
\bea
\mlrgi  = 7.6(8) \mev \quad , \quad \msrgi = 177(14) \mev \quad ,
\eea
where the renormalization group invariant quark mass is defined according to the convention usually adopted  in
perturbative calculations~\cite{francolub,chetyrkin,larin}, {\it i.e.}
\bea
m^{\rm RGI}_q = \lim_{\mu\to \infty} m_q(\mu)\left( \alpha_s(\mu)\right)^{- \gamma_m^{(0)}/\beta_0} \;,
\eea 
with $\gamma_m^{(0)}$ and $\beta_0$, being scheme independent~\footnote{Note that in Ref.~\cite{npm_sf}, another convention has been used:
\bea
m^{\rm RGI}_q = \lim_{\mu\to \infty} m_q(\mu)\left( {2 \beta_0 \over \pi}\alpha_s(\mu)\right)^{- \gamma_m^{(0)}/\beta_0} \;.
\eea
}.
 
Finally, the quark masses in the  $\overline {\rm MS}$-scheme  read
\bea
\label{MSmasses}
&& \hspace*{3.2cm}  {\scriptsize{\textsf{NLO}}} \quad  \; \;  {\scriptsize{\textsf{N$^{\sf 2}$LO}}}\quad   \; \;   {\scriptsize{\textsf{N$^{\sf 3}$LO}}} \quad \cr 
&& \hfill \cr
&& \mlmsbar (2\gev) = \left\{ 5.2(5);\ 4.9(5);\ 4.8(5) \right\}\ \mev
\quad ,\cr
&&\cr
&& \msmsbar (2\gev) = \left\{ 120(9);\  114(9);\ 111(9)\right\}\ \mev
\quad ,
\eea
where the numbers within the curly brackets are obtained after converting the 
RI-MOM results to the $\overline {\rm MS}$ one to ${\rm NLO}$, ${\rm N^2LO}$
and ${\rm N^3LO}$ accuracy, respectively. 
The details on anomalous dimensions and beta function are listed in the appendix.

We note that in most of the phenomenological applications, for example with QCD sum rules,
the theoretical expressions are only known to the NLO and, for consistency, the quark masses at the same accuracy should be used.

Similarly, we stress that lattice calculations of quark masses, in which the mass renormalization constants has been determined by using (one-loop) perturbation theory, should be compared with our NLO results of Eq.~(\ref{MSmasses}), since they have been derived at the same order of accuracy.

\par
Preliminary results obtained with the method discussed in this paper were already presented at the ``Lattice 99" Conference~\cite{vittorio}.
\section{Improved quark propagator}
\label{sec:improv}

\par
The general problem of improving gauge non-invariant, off-shell, 
correlation functions has been studied in Ref.~\cite{unpub}. Since this paper is still unpublished, in this section we discuss in some detail the non-perturbative improvement of the lattice quark propagator to $\Oa$.

\subsection{The subtracted quark propagator}
\par
The original idea of improvement~\cite{symanzik} 
(later developed in Ref.~\cite{luscher} for gauge theories) consists in adding, to both the action and operators, a 
complete set of higher-dimensional ({\sl ``irrelevant"})
operators, the coefficients of which are tuned as to cancel 
finite cut-off effects (to a desired order of lattice spacing). 
Specifically, the improvement of the Wilson action to $\Oa$ is 
achieved by adding a set of dimension-five operators~\cite{a1}, 
\bea
S = S_W + a \sum_{i=1}^{n} c_i \int d^4 x \ {\cal O}_i^{(d=5)} (x)\
,
\eea
allowed by  gauge invariance and discrete lattice symmetries, namely
\bea
&& \quad \quad \; {\cal O}_1\ = \ {i\over 4} \ \bar q\ \sigma_{\mu\nu}F_{\mu\nu}\ q\,,
 \quad {\cal O}_2\ =\ {1\over 2 g_0^2}\ m\ {\rm Tr} \bigl(  F_{\mu\nu}F_{\mu\nu} \bigr),
  \nonumber
  \\
  \noalign{\vskip1ex}
&& {\cal O}_3\ =\ m^2\ \bar q\ q\,, \quad 
 {\cal O}_4\ =\ m \ \bar q\ \bigl( \lrvec \Dslash + m_0 \bigr) q\,, \quad 
 {\cal O}_5 \ = \ \bar q\ \bigl( \lrvec \Dslash + m_0 \bigr)^2 q\, ,
  \label{ops}
\eea
where  $m$ is the bare  subtracted mass and $\Dslash + m_0$ the
bare Dirac operator appearing in $S_W$. The operators ${\cal O}_2$ and ${\cal O}_3$ can be reabsorbed in 
the definition of the bare strong coupling and the quark mass.
A major simplification comes from the restriction 
of  improvement to physical amplitudes, for which the equations 
of motion can be used. In this way, one is left with  one  (Clover) counterterm only,  ${\cal O}_1$, the coefficient of which was computed non-perturbatively in Ref.~\cite{a1}.

\par
The equations of motion cannot be used to improve   off-shell 
quantities (such as the  quark propagator): in this case, one  must also
consider the operators ${\cal O}_{4,5}$. As for  operators which are
BRST allowed but not gauge invariant, only the  BRST variation of 
$\overline{c}^a \partial_\mu A_\mu^a$ (where $\overline{c}^a$
and $A_\mu^a$ are the anti-ghost and gauge fields,
respectively) may contribute. This term, however,  can be 
absorbed into a redefinition of the gauge-fixing
parameters~\cite{unpub}. Besides the terms in the action considered above,
in Ref.~\cite{unpub} it has been also shown that, for the quark
propagator, there is another
operator, which is not BRST invariant but may contribute to 
off-shell correlation functions (because its presence is 
not excluded by Slavnov-Taylor identities). The effects of
${\cal O}_{4,5}$ and of this extra operator can be eliminated 
with a simple redefinition of the quark field:
\bea
\widehat q(x) = Z_q^{(0)\ -1/2} (1 + b_q ma) \left\{ 1\ +\ a \, \cqp \left( 
\Dslash + m_0 \right) 
\ +\ a \, \cngi \, \dslash \right\}\ q(x)\ .
\label{eq:qimp}
\eea

\par
We now discuss how the unknown coefficients $Z_q$ 
($Z_q^{-1/2} = Z_q^{(0)\ -1/2}(1 + b_q ma)$), $\cqp$ (corresponding to the coefficient of the operator ${\cal O}_5$) and $\cngi$, present in 
Eq.~(\ref{eq:qimp}), can be determined from the analysis of the lattice 
bare quark propagator, $S_L(p)$. 

\par
From Eq.~(\ref{eq:qimp}), it follows that the relation between $S_L(p)$ 
and the improved, renormalized quark propagator, $\widehat 
S(p)$, constructed in terms of the quark fields, $q$ and $\widehat q$ 
respectively, has the form:
\be
S_L(p) = \left( 1 -2 a \, \cngi \, i \slash{p} \right) Z_q \widehat S(p) 
- 2 a \, \cqp\ \; .
\label{eq:simp}
\ee
In this equation, it is convenient to express the renormalized quark 
propagator, $\widehat S(p)$, in terms of the two invariant scalar form 
factors, $\suno (p^2)$ and $\sdue (p^2)$, defined in Eq.~(\ref{eq:sigma12}).
For further use, we remark that at large $p^2$,
up to power-suppressed ($\sim 1/p^2$) and logarithmic corrections,
$\suno (p^2) \simeq 1$ and $\sdue (p^2) \simeq m$, where $m$ is the 
renormalized quark mass. 
After substituting (\ref{eq:sigma12}) into (\ref{eq:simp}), one finds:
\bea
&&\sigma_{1L}(p^2) = \frac{1}{12} \, {\rm Tr} 
\biggl[ -i \slash{p} S_L(p) \biggr]= 
Z_q \biggl( \suno (p^2) - 2 a \, \cngi \,
 \sdue (p^2) \biggr) \label{eq:tr1}
\\
&&\hfill \cr
&&\hfill \cr
&&\frac{\sigma_{2L}(p^2)}{p^2} = \frac{1}{12} \, {\rm Tr} \biggl[ S_L(p)
\biggr] 
= -2 a \, \cqp + 2 a \, \cngi \, Z_q \suno (p^2) + Z_q \frac{\sdue(p^2)}{p^2}
\label{eq:tr2}
\eea
where $\sigma_{1L,2L}(p^2)$ are the analog of $\widehat \sigma_{1,2}(p^2)$
for  the lattice bare propagator. 
\par 
Using  Eqs.~(\ref{eq:tr1}) and (\ref{eq:tr2}), the coefficients $\cqp$  
and $Z_q \cngi$ can be determined as follows: at large $p^2$ and in the chiral limit, since
$\sdue \sim 1/p^2 \to 0$, it is in principle possible to separate $\cqp$ 
and $Z_q \cngi$ using the $p^2$ dependence of $\suno$; 
the overall
renormalization constant $Z_q$, including its $\Oa$ mass dependence, can
then be determined by combining Eq.~(\ref{eq:tr1}) with the renormalization
condition~(\ref{eq:rcs}). In practice, however, this procedure is very 
difficult to implement. The reason is that the logarithmic $p^2$-dependence of 
$\suno (p^2)$, entering the right-hand side of Eq.~(\ref{eq:tr2}), is very
mild. This is especially true in the Landau gauge, where it starts at order 
$\alpha_s^2$ in perturbation theory. Thus, it is very hard (if not impossible) 
to disentangle the contributions coming from the two coefficients~\footnote{A promising way to compute 
separately $\cqp$ and $\cngi$ non-perturbatively is from the study of the quark-gluon vertex.}, $\cqp$ and 
$\cngi$.

Let us return to the large-$p^2$ behaviour of the lattice quark propagator,
 in the limit where  power and logarithmic corrections can be neglected. 
In this limit, Eqs.~(\ref{eq:tr1}) and (\ref{eq:tr2}) become:
\be
\sigma_{1L}(p^2) \simeq Z_q \biggl( 1 - 2 a \, \cngi \, m \biggr) \equiv 
\widetilde Z_q\,,
\label{eq:tr1nl}
\ee
and 
\be
\frac{\sigma_{2L}(p^2)}{p^2} \simeq -2 a \, \cqp + 2 a \, \cngi \, Z_q
\equiv 
-2 a \, \widetilde c_q^{\,\prime}\,.
\label{eq:tr2nl}
\ee
In the large-$p^2$ region, by using Eq.~(\ref{eq:tr1nl}), it is
then possible to determine an {\sl ``effective"}  renormalization
constant,
$\widetilde Z_q$, which reduces to $Z_q$ in the chiral limit. Moreover, in
terms of the coefficient $\widetilde c_q^{\,\prime}$, computed through 
Eq.~(\ref{eq:tr2nl}), we can define a {\sl ``subtracted"} quark propagator, 
$\widetilde S(p)$, as:
\bea
\widetilde S(p) = \widetilde Z_q ^{-1} \biggl(  S_L(p) + 2 a \, \widetilde 
c_q^{\,\prime} \biggr) \,.
\label{eq:seff}
\eea 
If the coefficient $\cngi$ were equal to zero, the
subtracted propagator $\widetilde S(p)$ would correspond to the improved,
renormalized quark propagator, $\widehat S(p)$, as can be seen from 
 Eq.~(\ref{eq:simp}). 
In the presence of $\cngi$,  however,  $\widetilde S(p)$
and $\widehat S(p)$  differ by terms of  ${\cal O}(\cngi\ am)$, up to
(small) logarithmic corrections. 
We conclude that, by following the procedure outlined above, we are able
to exactly improve the quark propagator in the chiral limit. Out of the
chiral  limit, since $\cngi$ is of ${\cal O}(g_0^2)$ in perturbation
theory,  the propagator is affected by ${\cal O}(g_0^2 am)$
discretization errors. In the range of quark masses considered in this
paper, these terms are expected to be smaller than other statistical and 
systematic uncertainties. They may be important, however, in the calculation of  heavy quark masses.

\subsection{Practical implementation}

\begin{figure}[t!]
\begin{center}
\begin{tabular}{@{\hspace{-1.0cm}}c c c}
\epsfxsize9.0cm\epsffile{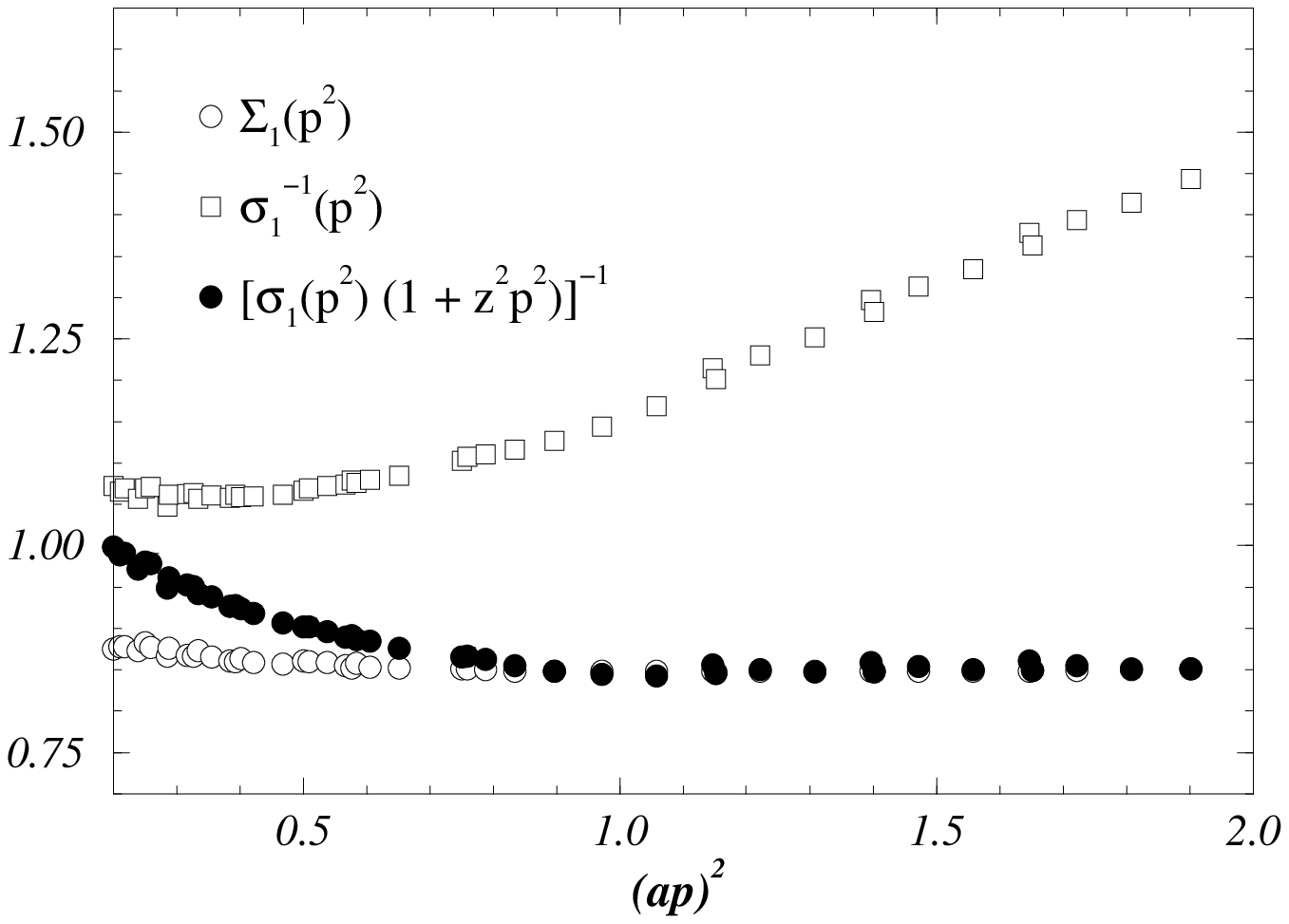} &\hspace*{-0.5cm} & \epsfxsize9.0cm\epsffile{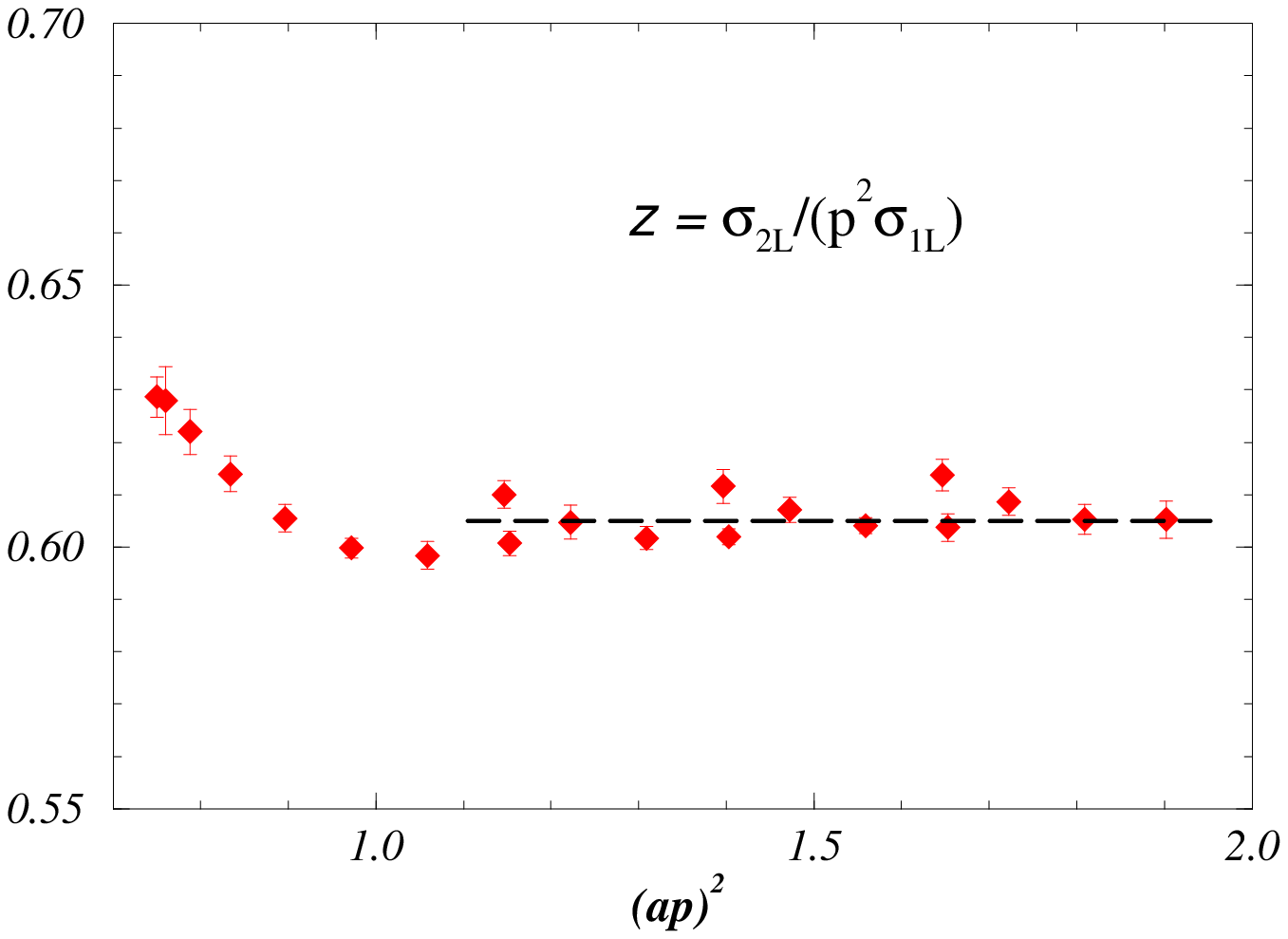}  \\
\end{tabular}

\caption{\label{PROP1}{\sl The left figure shows the effect of the correction due to the factor
$(1 + z^2 p^2)$. Empty symbols denote the quantities extracted directly from the lattice quark propagator. The filled circles denote the effect of the correction. Note that the error bars are smaller than the used symbols. The value of the constant $z$ as obtained from a fit to our data in the large $p^2$-region is shown in the right figure. Both figures correspond to  $\kappa=0.1344$.}}
\end{center}
\end{figure}

Let us now discuss how  the improvement procedure for the lattice quark 
propagator works in practice. 
As a preliminary (and instructive) step, we consider the
inverse unsubtracted lattice propagator  expressed  in terms of the
usual form factors, 
$\Sigma_{1L}$ and $\Sigma_{2L}$:
\be
S_L^{-1}(p)  = -i \slash{p} \, \Sigma_{1L} (p^2) + \Sigma_{2L} (p^2)\,.
\label{eq:sigma12b}
\ee
$\Sigma_{1L}$ is special in that its  $p^2$-behaviour is protected 
by the VWI, $\Sigma_{1L}(p^2)\sim$ const.$=Z_V$, as can be seen in
Fig.~\ref{PROP1} (empty circles)~\footnote{Throughout this paper, we adopt a continuum notation
in
which $a p_\mu$ stands for $\sin(a p_\mu)$. Thus, for instance, $a^2 p^2$ corresponds to $\sum _\mu \sin^2 (a p_\mu)$ and $a \slash p$ is equal to
$\sum_\mu \gamma_\mu \sin (a p_\mu)$.}. By using Eqs.~(\ref{eq:sigma12}) and
(\ref{eq:tr2}),  in the large-$p^2$ limit one finds
\bea
\sigma_{1L}(p^2) \ = \ \Sigma_{1L}^{-1}(p^2) \, \left[\ 1 \ +\ \left(
\dfrac{\sigma_{2L}(p^2)}{p^2 \sigma_{1L}(p^2)} \right)^2 p^2 \
\right]^{-1}\ \to \ 
\Sigma_{1L}^{-1}(p^2) \, \left[\ 1\ +\  {\cal O}(a^2 p^2)\ \right]\; .
\label{eq:guido}
\eea
The effect of the ${\cal O}(a^2)$ term in Eq.~(\ref{eq:guido}) is important at large $p^2$, as shown in Fig.~\ref{PROP1}: $\Sigma_{1L}(p^2)$ (empty circles) is indeed flat, whereas $\sigma_{1L}^{-1}(p^2)$ (empty squares) is not. 
The reason is the presence of the contact term 
$\widetilde c_q^{\,\prime}$ in the factor 
\bea 1+ z^2 p^2 \  \equiv \ 
 1+ \left({\sigma_{2L}(p^2)\over p^2 \sigma_{1L}(p^2)}\right)^2 p^2 \ \to \  1 \ +\  4 {\ \widetilde c_q^{\prime 2} }  a^2 p^2  \ ,
\eea
at large $p^2$. For $\widetilde c_q^{\,\prime}=0$, instead, 
the factor $ 1+ z^2 p^2 \sim 1+  m^2/p^2 \to 1 $.
Thus the dangerous ${\cal O}(a^2)$ terms of 
Eq.~(\ref{eq:guido}) jeopardizes the $p^2$ behaviour 
that $\sigma_{1}$ should have in the continuum.

The following procedure has been adopted to get rid of the  ${\cal O}(a^2)$ corrections induced by the contact terms~\footnote{ Although not
mentioned before, it is clear that there are other ${\cal O}(a^2)$
effects,  besides those  induced by the ${\cal O}(a)$ contact term, resulting in the factor $(1 + z^2 p^2)$.
These, however, are found to be much smaller, see below.}.  
Motivated by Eq.~(\ref{eq:guido}), we first multiply
the original lattice propagator by the overall factor, $\left( 1 + z^2 p^2 \right)$.
Since the action
is only $\Oa$-improved, this subtraction is formally irrelevant. 
The multiplication, however, removes the dependence of
$\sigma_{1L}(p^2)$ on $p^2$, coming from the $\Oaa$ effect discussed above.
This can be seen in Fig.~\ref{PROP1} by comparing the squares and filled
circles.  It is important to add that $z$ (for each $\kappa$)
has been fixed by fitting to a constant the ratio
\bea
\label{smallz}
z =  {1 \over  \; p^2 \; } {\sigma_{2L}\over \sigma_{1L}} \ ,
\eea
in the large $p^2$-region.
The data and the fit are displayed in Fig.~\ref{PROP1} for $\kappa= 0.1344$.
The numerical values for $z$ will be given in the next section.
The multiplication by $(1 + z^2 p^2)$ flattens the $p^2$ dependence of both $\sigma_{1L}$ and $\sigma_{2L}$. This effect is particularly pronounced for $\sigma_{1L}$, as can be seen in Fig.~\ref{PROP1}. 


We now describe the procedure followed to remove the constant contact term $c_q^{\,\prime}$, in Eq.~(\ref{eq:seff}).
After the multiplication of the propagator by $(1 + z^2 p^2)$, 
we fit $\sigma_{2L}(p^2)$ to the form expected from the OPE, namely (up to logarithmic corrections)
\bea
{\sigma_{2L}(p^2)\over p^2} = {\cal A} + \frac{{\cal B}}{a^2 p^2} + {\cal
C} a^2 p^2\, ,
\label{eq:s2fit}
\eea
in the region of large Euclidean momenta where the OPE applies.
We have used the interval $0.5 \le a^2 p^2 \le 2.0$, which at $\beta=6.2$
corresponds to  $2\gev^2 \leq p^2 \leq 15\gev^2$ in physical units.
By comparison with Eq.~(\ref{eq:tr2nl}), the coefficient ${\cal A}$ 
is evidently ${\cal A}= -2 a \,\widetilde c_q^{\,\prime}$. 
From the fit of our data to Eq.~(\ref{eq:s2fit}), we obtain 
 ${\cal A}=0.617(11)$, to be compared to ${\cal A}^{\rm (BPT)}=0.573$, 
as computed in  one-loop (boosted) perturbation
theory~\cite{capitani}~\footnote{The perturbative calculation of
 $\widetilde c_q^{\,\prime}$ indicates  explicitly that this coefficient
is  gauge-dependent.}. According to Eq.~(\ref{eq:master}), the
parameter ${\cal B}$ is proportional to the quark mass, so that it is 
expected to vanish in the chiral limit. The value that we obtain,  ${\cal B}=0.003(6)$, is well consistent with expectations. It
clearly demonstrates that the contribution of the term proportional to the quark condensate is negligible in the  range of momenta chosen for the fit. 
This point was recently questioned 
in Ref.~\cite{alain}.
From our fit
this contribution appears to be completely negligible for $p^2 \gtrsim 3 \gev^2$,
as expected from OPE, if $\langle \bar q q \rangle \sim \Lambda_{QCD}^3$. 
Finally, we also obtain ${\cal C} = 0.022(4)$, for
the parameter which contains the information on residual $\Oaa$ effects. Note that without the correcting factor $(1+z^2 p^2)$ we would
have obtained ${\cal C}=-0.070(3)$, larger than the above results. This shows that the residual $\Oaa$ effects are smaller than those induced by the contact terms. 

\par
To summarize, we subtract the unwanted discretization effects using
\bea 
\label{subtr}
\widetilde S (p) = \frac{i \slash{p}}{p^2} \, \widetilde \sigma_1 (p^2) + \frac{\widetilde \sigma_2 (p^2)}{p^2} = S_L(p) \biggl( 1 +  z^2 p^2\biggr) - {\cal A} - {\cal C} a^2 p^2\ .
\eea
\begin{figure}[h!]
\begin{center}
\begin{tabular}{@{\hspace{-1.0cm}}c c c}
\hfill  & \epsfxsize9.0cm\epsffile{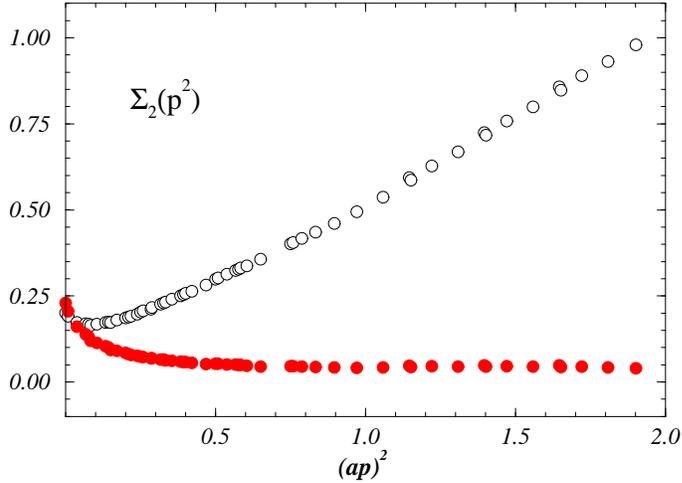}&  \hfill  \\
\end{tabular}
\caption{\label{PROP2}{\sl 
Empty symbols denote $\Sigma_{2L}(p^2)$, extracted directly from the lattice quark propagator. The filled symbols correspond to the subtracted form factor $\widetilde \Sigma_{2L}(p^2)$. Illustrated is the case with $\kappa=0.1344$.}}
\end{center}
\end{figure}

The resulting improved propagator, 
$\widetilde S(p)$, exhibits a good chiral behaviour at large $p^2$
and its inverse has small ${\cal O}(a^2)$ corrections.
$\Sigma_{2}(p^2)$ is expected to be a slowly varying function
of the momentum at large $p^2$, with a logarithmic $p^2$-dependence 
governed by the quark mass anomalous dimension. 
As shown in Fig.~\ref{PROP2}, the bare form factor ($\Sigma_{2L}(p^2)$) exhibits, instead, a strong linear dependence in
$(pa)^2$, induced by the $\Oa$ contact term proportional to $\widetilde c_q^{\prime}$. This effect disappears in the subtracted form factor $\widetilde \Sigma_{2L}(p^2)$, defined from $\widetilde S^{-1} (p)$, as shown in the same plot.

\begin{figure}[t!]
\begin{center}
\begin{tabular}{c c c}
\hfill &\epsfxsize11.0cm\epsffile{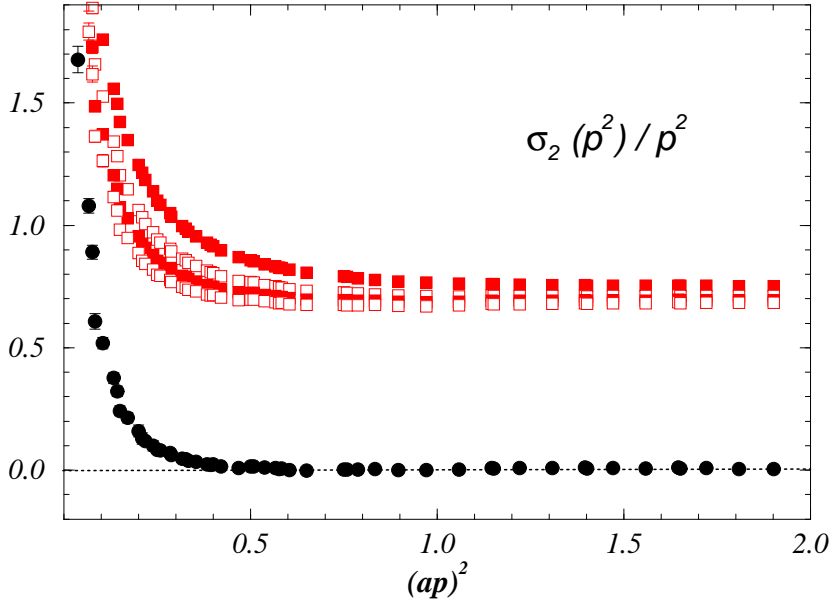}&\hfill  
\end{tabular}

\caption{\label{Traces}{\sl The effect of subtraction of the contact term, $\widetilde c_q^\prime$, from the scalar part of the quark propagator. Squares denote $\sigma_2/p^2$, as obtained 
from unsubtracted (albeit corrected by the $(1 + z^2 p^2)$ factor) propagators corresponding to the four values of the hopping parameters used in this study.
The filled symbols denote $\widetilde \sigma_2/p^2$, obtained after the subtraction of the term due to $\widetilde c_q^\prime$ (see Eq.~{\rm (\ref{subtr})} ), and extrapolated to the chiral limit.}}
\end{center}
\end{figure}

As a further confirmation of the effectiveness of the subtraction, we show $(1+z^2 p^2) \sigma_{2L} (p^2)/p^2$ in Fig.~\ref{Traces} for different values of $\kappa$. 
At large $p^2$ and for all the values of the quark masses,
this quantity has a good plateau
corresponding, up to further $\Oaa$ corrections, to the contact term $\sim
\widetilde c_q^{\,\prime}$ that we want to subtract. In the same figure, we also show
$\widetilde \sigma_2(p^2)/p^2$, after the subtraction and extrapolated to the
chiral limit (filled circles). This curve
demonstrates the accuracy of the subtraction procedure, since in this limit we expect $\widetilde \sigma_2(p^2)/p^2 \sim \langle \bar q q \rangle
/p^4$. 

\section{Numerical details and physical results}
\label{sec:lattice}

\par
In this section we briefly recall some elements of the lattice calculation which are explained in great detail in Ref.~\cite{light}, and present our physical results. These results are obtained on a sample of 200 quenched gauge fields configurations, on a $24^3 \times 64$ lattice and at $\beta=6.2$. 
The value of the inverse lattice spacing, $a^{-1} = 2.72(11)$~GeV, is obtained from the
$m_{K^*}$ mass. The quark propagators have been computed by 
using the non-perturbatively $\Oa$-improved Wilson 
action, for four different values of the light quark masses which correspond to the following set of the hopping parameters:
\bea
\label{kappas}
\kappa \in \left\{ 0.1352, 0.1349, 0.1344 , 0.1333 \right\}\ .
\eea
\noindent 
All other details, concerning the analysis of the light hadron spectrum, 
can be found in Ref.~\cite{light} where a subset of $100$ configuration was analyzed. 
%
\begin{figure}[t!]
\begin{center}
\begin{tabular}{c c c}
\hfill &\epsfxsize11.0cm\epsffile{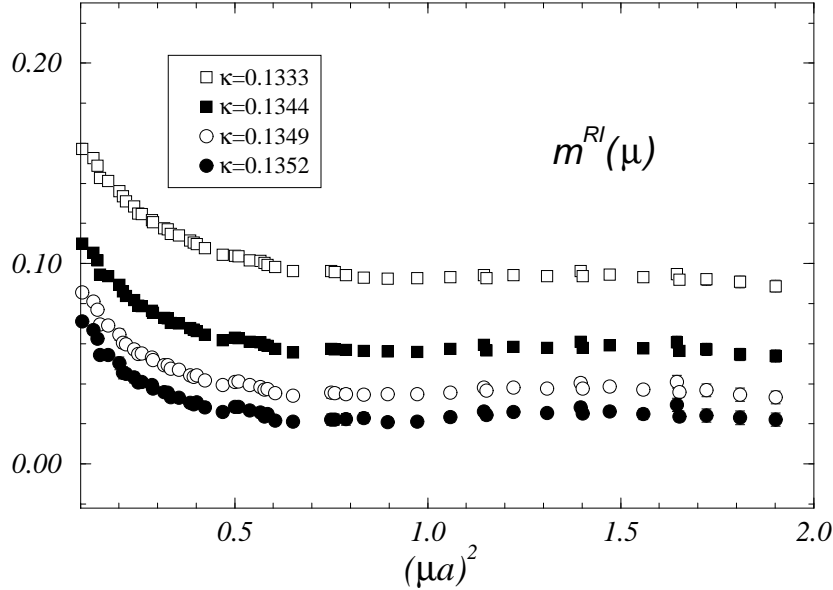}&\hfill  
\end{tabular}
\caption{\label{QMASS}{\sl The lattice quark masses corresponding to the indicated $\kappa$ values, in the RI-MOM scheme.
They have been obtained using Eq.~{\rm (\ref{mri})}. The error bars are not visible, because they are smaller than the symbols used in this figure.}}
\end{center}
\end{figure}

\par
As discussed in the previous section, Eq.~(\ref{subtr}) involves the determination of the 
constant $z$. From the fit of our data to~(\ref{smallz}) in the interval  
$1.1 \leq (ap)^2 \leq 2$, we obtain:
\be
z = \left\{ 0.581(3),\ 0.591(2),\ 0.605(3),\ 0.639(2) \right\}\,,
\ee
in decreasing order w.r.t. the $\kappa$-parameter.
With the value of $z$ at hand, we follow the subtraction procedure
described in the previous section and obtain the renormalized propagator,
from which the quark masses can be derived.
We now present the numerical results for $Z_q$ and for the quark masses in different schemes.

\subsection{The quark field renormalization constant $Z_q$}

We fit the form factor $\widetilde \Sigma_{1L}(p^2)$ to a constant, in the large momentum region.
According to Eq.~(\ref{eq:rcs}), the value of that constant corresponds, up to 
tiny  ${\cal O}(\alpha_s^2)$ corrections~\cite{francolub,chetyrkin}, to the value of the
quark field renormalization constant, $Z_q$, in the RI-MOM scheme. 
From a fit in the region of large $(ap)^2\in [1  ,1.8]$ 
(corresponding
to $7.4\lesssim p^2 \lesssim 13.2~{\rm GeV}^2$ in physical units), we get 
\be
\widetilde Z_q = \left\{ 0.849(2),\ 0.847(2),\ 0.846(3),\ 0.839(2) \right\}\,.
\ee
These values extrapolated (linearly) to the chiral limit
give~\footnote{The value of $\kappa_{crit}$ is $0.135855(19)$.}:
\be
Z_q^{\rm (0)} (2\ \gev) = 0.853(3)\,,
\label{eq:zq}
\ee
which is to be compared with $Z_q^{\rm (0)} = 0.875$,
 from one-loop (boosted) perturbation theory~\cite{capitani}. Note that
the quark mass dependence of $\Sigma_{1L}(p^2)$, 
which according to Eq.~(\ref{eq:tr1nl}) 
comes from both the mass dependent term in $Z_q$ and the term
proportional to $\cngi$, is rather weak.


\subsection{Extracting the physical quark mass}
 
From the subtracted inverse propagator, $\widetilde S^{-1}(p)$, we get the renormalized quark mass in the RI-MOM scheme which, according to Eq.~(\ref{eq:master}), corresponds at large $p^2$ to  
\bea
\label{mri}
m^{\rm RI}(\mu)\ =\ \left. {\widetilde \Sigma_2 (p^2)\over \widetilde \Sigma_1 (p^2)} \; \right|_{p^2 = \mu^2} \ .
\eea
This quantity  is unaffected by the correcting factor 
$(1 + z^2 p^2)$. In Fig.~\ref{QMASS}, we display $m^{\rm RI}(\mu)$ for different values of $\kappa$.  
The numerical value of $m^{\rm RI}(\mu)$ can be read off directly from Fig.~\ref{QMASS}. Note that $m^{\rm RI}(\mu)$ is derived in a completely non-perturbative way. 

In practice, to reduce the statistical fluctuations, we extract $\mri(\mu_0)$, from a fit in the interval $1.1 \leq (\mu a)^2 \leq 1.8$, corresponding to $7.5 \gev^2 \lesssim \mu^2 \lesssim  13 \gev^2$, by using 
\bea
\label{fittt}
\mri(\mu) = {c^\ri (\mu)\over c^\ri (\mu_0)}\ \mri(\mu_0)\ , 
\eea
with $c^\ri (\mu)$ computed in perturbation theory. We have chosen $(\mu_0 a)^2=1.45$, in the middle of the fitting interval, corresponding to $\mu_0=3.28$~GeV.
With this procedure the error induced by the use of perturbation theory is negligible, namely we find that the differences between NLO and N$^3$LO are less than $1\ \%$. 
The reason is that the expression in Eq.~(\ref{fittt}) depends on the ratio of $c^\ri$'s evaluated at different but close scales.

The numerical values of $c^{\rm RI}(\mu)$ has been obtained
by using the quenched expression for $\alpha_s(\mu)$
with $\Lambda_{\rm QCD}= 318$~MeV~\cite{propag}. We checked that by using $\Lambda_{\rm QCD}= 238$~MeV, as found in Ref.~\cite{npm_sf}, the central value
of the masses is increased by less than $1 \%$.

To compute the physical values of the light and the strange 
quark masses, we invoke the procedure described in Ref.~\cite{mq_noi,light}. The masses are fitted as quadratic functions of the squared pseudoscalar meson masses. 
By using the method of ``physical lattice planes''~\cite{mq_ape96}, we fix the average up-down and the strange quark masses, from the $\pi$ and $K$ meson respectively~\footnote{Note that consistent results are obtained when the $\phi$-meson is used to extract the strange quark mass.}.  
The results are the following:
\bea
\mlri(3.28\ \gev) = 5.1 (5)\ \mev
\; ,\hspace*{1.3cm}
\msri(3.28\ \gev) = 118 (9)\ \mev \; .
\eea
Since it is costumary to give the quark masses at the renormalization scale $\mu=2$~GeV, we use again Eq.~(\ref{fittt}) to rescale $\mri(3.28 \gev)$ to $\mri(2 \gev)$. This time, we have to run the mass to a lower scale than the one used in the fitting procedure ($7.5 \gev^2 \lesssim \mu^2 \lesssim  13 \gev^2$). For this reason the uncertainty due to higher orders is larger than before, namely of the order of $4\ \%$.
The results are those given in Eq.~(\ref{RImasses}).
We note, in passing, that there is no reason, if not for comparison with other calculations, to evolve the masses down to $\mu \ = 2 \gev$. Indeed, it would be much more convenient to work at a scale $\mu \gtrsim 3 \gev$ where the uncertainty induced by higher order perturbative corrections is negligible.

\begin{figure}[h!]
\begin{center}
\begin{tabular}{c c c}
\hfill & \epsfxsize11.0cm\epsffile{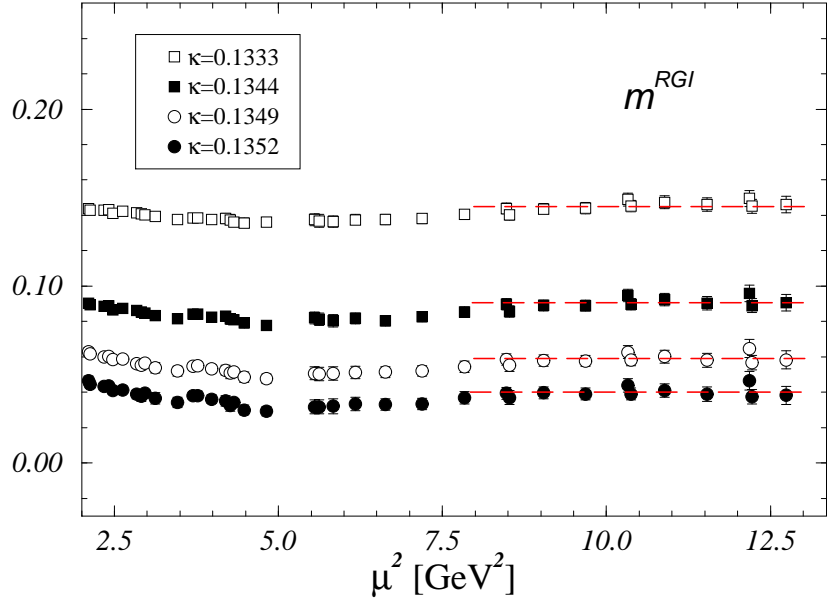}&\hfill  
\end{tabular}

\caption{\label{QMASSRGI}{\sl 
Renormalization group invariant quark masses, obtained after dividing out the scale depending part in the RI-MOM scheme to $\rm N^3LO$ accuracy, $c^{\rm RI}(\mu)$, from the quark masses depicted in Fig.~{\rm \ref{QMASS}}. Dashed lines correspond to the fitting interval used in~{\rm (\ref{fittt})}.}}
\end{center}
\end{figure}

We now illustrate the procedure adopted to obtain the quark masses in different schemes. The renormalization group invariant quark mass, which is a scheme and scale independent quantity, is related to $m^{\rm RI}(\mu)$ by the expression 
\bea
\label{RGI}
m^{\rm RGI}_q \ =\ {\mqri (\mu) \over c^{\rm RI}(\mu)}\,,
\eea
whereas for the $\msb$ mass we have
\bea
\label{MASSMS}
\mqmsbar (\mu)\ =\  c^{\msb}(\mu)\ m^{\rm RGI}_q\ =\ { c^{\msb}(\mu)\over c^{\rm RI}(\mu)}\ \mqri (\mu)  \ . 
\eea
The resulting values of $m^{\rm RGI}_q$ should be flat in a large range of $\mu^2$, as confirmed by the data shown in Fig.~\ref{QMASSRGI}. 
By using~(\ref{RGI}), we obtain the following results
\bea
\label{resRGI}
&& \hspace*{2.2cm}  {\scriptsize{\textsf{NLO}}} \quad  \; \;  {\scriptsize{\textsf{N$^{\sf 2}$LO}}}\quad   \; \;   {\scriptsize{\textsf{N$^{\sf 3}$LO}}} \quad \cr 
&& \hfill \cr
&& \mlrgi = \left\{ \ 8.5(9);\; 7.8(8);\; 7.6(8) \right\}\ \mev
\quad ,\cr
&&\cr
&& \msrgi = \left\{ 196(15);\  182(14);\ 177(14)\right\}\ \mev
\quad .
\eea
In the $\msb$ case, from Eq.~(\ref{MASSMS}) we get the results quoted in Eq.~(\ref{MSmasses}).

In the calculation of $\mqrgi$ and $\mqmsbar(\mu)$, we have used
the physical value of $\alpha_s(\mu)$, corresponding to $\alpha_s(M_Z)=0.118$~\cite{PDG}, computed with the appropriate number of active flavors, {\it e.g.} $n_f=4$ at $2$~GeV. This choice can be justified by assuming that the masses in Eq.~(\ref{RImasses}) are the physical ones, up to some unknown quenching errors. We checked, however, that the results by using the quenched $\alpha_s$ with $\Lambda_{\rm QCD}=318$~MeV, would be different by less than $2$~\% in all the cases considered in this paper.

\section*{Conclusion}

We have applied a new method to compute the renormalized quark masses from
the lattice quark propagator using the OPE. We have discussed the subtleties
related to the improvement of the propagator and especially the troubles arising from the presence of contact terms. Some of these problems 
could be avoided by working with $S_L(x)$ instead than $S_L(p)$. Feasibility studies are underway. The main results, given in the introduction, are well compatible with values of the masses obtained by standard lattice methods. They  are also in very good agreement with the recent result of Ref.~\cite{pichprades}, $m_s^{\overline{\rm MS}}( 2\ \gev) = 114 \pm 24\ {\mev}$, obtained by using the model independent QCD sum rule analysis at a ${\rm N^3LO}$ (see also Ref.~\cite{narison}).

\section*{Acknowledgements}
We thank Massimo Testa and Konstantin Chetyrkin for illuminating and valuable discussions. We are also grateful to Chris Dawson, Alain Le Yaouanc, Carlotta Pittori,
Steve Sharpe and Gian Carlo Rossi for interesting discussions on the implementation of the method presented in this paper. V.L. and G.M. acknowledge the M.U.R.S.T. and the INFN for partial support. D.B. thanks the INFN for financial support.

\vspace{1.5cm}

\section*{Appendix}

In this appendix we list the formulae which have been used to compute the perturbative scale dependence of the quark masses.

 The effective QCD coupling is governed by the $\beta$-function which is known to 4-loops:
\be
\mu^2 {d\over d \mu^2} \left({\alpha_s(\mu)\over \pi}\right) =
-  \sum_{n=0}^3 \beta_n   \left({\alpha_s(\mu)\over \pi}\right)^{n+2} +\ {\cal O}(\alpha_s^6(\mu)) \,,
\ee
where the coefficients are~\cite{vermas}:
\bea
&& \beta_0 = \frac{\large 1}{4}\left(11 - {2\over 3}n_f\right) \,,   \quad  \quad \beta_1 = { 1\over 16} \left(102 - {38 \over 3} n_f\right)\,,\cr
{\phantom{\huge{l}}}\raisebox{-.1cm}{\phantom{\Huge{j}}}
\hfill  &&\hfill \cr
&& \beta_2^{\overline{\rm MS}} = { 1\over 64} \left( \frac{2857}{2} - {5033 \over 18} n_f + {325 \over 54} n_f^2\right)\,, \cr
\hfill  &&\hfill \cr
\hfill  &&\hfill \cr
 &&  \beta_3^{\overline{\rm MS}} = { 1\over 256} \left[ \frac{149753}{6} - 3564 \zeta(3) - \left(\frac{1078361}{162} + {6508 \over 27} \zeta(3)\right) n_f \ + \right.  \cr
 &&  \hspace*{2cm}\left. \left(\frac{50065}{162} + {6472 \over 81} \zeta(3)\right) n_f^2 + \frac{1093}{729} n_f^3 \right]
\,.
\eea
The  coefficients of the mass anomalous dimension which describes the running 
of the quark mass,
\bea
\mu^2 {d\over d \mu^2}m_q(\mu) = - m(\mu) \sum_{n=0}^3 \gamma_m^{(n)}   \left({\alpha_s(\mu)\over \pi}\right)^{n+1} + \ {\cal O}(\alpha_s^5(\mu)) \,,
\eea
are also known up to four loops in both RI~\cite{chetyrkin} and $\overline{\rm MS}$~\cite{larin} schemes. We list them all:
\bea
\gamma_m^{(0)} = 1 \;, \nonumber \eea
\vspace*{-6mm}
\bea
 \left(\gamma_m^{(1)}\right)^{\overline{\rm MS}} = { 1\over 16} \left({202 \over 3} - {20 \over 9} n_f\right)\,, && 
\left(\gamma_m^{(1)}\right)^{\rm RI} = { 1\over 16} \left(126  - {52 \over 9} n_f\right)\,,\nonumber \eea
\vspace*{-6mm}
\bea
&&\left(\gamma_m^{(2)}\right)^{\overline{\rm MS}} = { 1\over 64} \left[ 1249 - \left( {2216 \over 27} + {160 \over 3} \zeta(3) \right) n_f - {140 \over 81} n_f^2\right]\,,\cr
 && \hfill \cr
 && \hfill \cr
 && \left(\gamma_m^{(2)}\right)^{\rm RI} =   { 1\over 64} \left[ {20911 \over 3} - {3344 \over 3}\zeta(3) - \left( {18386 \over 27} - {128 \over 9} \zeta(3) \right) n_f + {928 \over 81} n_f^2 \right] \,,\cr
&& \hfill  \cr
&& \hfill  \cr
&& \left(\gamma_m^{(3)}\right)^{\overline{\rm MS}} =  \frac{1}{256} \left[ \frac{4603055}{162} + \frac{135680}{27}\zeta(3) - 8800 \zeta(5)\right.
- \cr
&&\hspace*{3cm} \left( \frac{91723}{27} + \frac{34192}{9} \zeta(3) - 880 \zeta(4) - \frac{18400}{9} \zeta(5) \right) n_f\ + \cr
&&\hspace*{3cm} \left.
 \left( \frac{5242}{243} + \frac{800}{9} \zeta(3) - \frac{160}{3} \zeta(4) \right) n_f^2
- \left( \frac{332}{243} - \frac{64}{27} \zeta(3) \right) n_f^3
\right] \,,\cr
&& \hfill  \cr
&& \hfill  \cr
&&\left(\gamma_m^{(3)}\right)^{\rm RI}  =  \frac{1}{256} \left[ \frac{300665987}{648} - \frac{15000871}{108}\zeta(3) + {6160\over 3} \zeta(5)\right.
- \cr
&& \hspace*{3cm}\left( \frac{7535473}{108} - \frac{627127}{54} \zeta(3) - {4160\over 3} \zeta(5) \right) n_f\ 
+ \cr
&&\hspace*{3cm}\left. \left( \frac{670948}{243} - \frac{6416}{27} \zeta(3) \right) n_f^2
-  \frac{18832}{719} n_f^3
\right]\,.
\eea
The corresponding evolution part in the running quark masses are:
\bea
&&c^{\rm RI}(\mu) = \alpha_s(\mu)^{4/11} \biggl\{ 1 + {489\over 242} \left({\alpha_s(\mu)\over \pi}\right) + \left[ {25335863 \over 1405536} - {19\over 6}\zeta(3) \right]\left({\alpha_s(\mu)\over \pi}\right)^2 +  \nonumber\\
&& \hspace*{2.7cm}\left[ {\frac{48247704573745}{220410533376}} - 
   {\frac{170324909}{2509056}}\zeta(3) + 
   {\frac{35}{36}}\zeta(5)\right]\left({\alpha_s(\mu)\over \pi}\right)^3
\biggr\}\ \ ;
\eea
\bea
&&c^{\overline{\rm MS}}(\mu) = \alpha_s(\mu)^{4/11} \biggl\{ 1 + {499\over 726} \left({\alpha_s(\mu)\over \pi}\right) +  
{\frac{6375961}{4216608}} 
\left({\alpha_s(\mu)\over \pi}\right)^2 +  \nonumber\\
&& \hspace*{2.7cm}\left[ 
  {\frac{344717507317}{55102633344}} + 
   {\frac{6293}{3564}}\zeta(3) - 
   {\frac{25}{6}\zeta(5)}\right]\left({\alpha_s(\mu)\over \pi}\right)^3
\biggr\}\ \ ,
\eea
for $n_f=0$, and 
\bea
&&c^{\rm RI}(\mu) = \alpha_s(\mu)^{12/25} 
\biggl\{ 1 + {8803\over 3750} \left({\alpha_s(\mu)\over \pi}\right) + \left[ {5679460183 \over 337500000} - {119\over 30}\zeta(3) \right]\left({\alpha_s(\mu)\over \pi}\right)^2 +  \nonumber\\
&& \hspace*{2.3cm}\left[ 
{\frac{14533180260067051}{91125000000000}} - 
   {\frac{1437607219}{21600000}}\zeta(3) + 
   {\frac{19}{4}}\zeta(5)\right]\left({\alpha_s(\mu)\over \pi}\right)^3
\biggr\}\ \ ;
\eea
\bea
&&c^{\overline{\rm MS}}(\mu) = \alpha_s(\mu)^{12/25} 
\biggl\{ 1 + {3803\over 3750} \left({\alpha_s(\mu)\over \pi}\right) + \left[ 
{\frac{793412683}{337500000}} - {\frac{4}{5}}\zeta(3)
\right]\left({\alpha_s(\mu)\over \pi}\right)^2 +  \nonumber\\
&& \hspace*{2.cm}\left[ 
 {\frac{57222640693973}{7593750000000}} - 
   {\frac{2202791}{337500}}\zeta(3) + 
   {\frac{5}{3}}\zeta(4) - 
 {\frac{7}{18}\zeta(5)}\right]\left({\alpha_s(\mu)\over \pi}\right)^3
\biggr\}\ ,
\eea
for $n_f=4$.

\end{document}